\documentclass[a4paper,12pt]{article}
\usepackage{epsfig}
\date{\today}
\begin{document}

\renewcommand{\thefootnote}{\fnsymbol{footnote}}

\rightline{SUSX-TH/01-036}
\vskip 1cm
\begin{center}
{\bf \large{  The CKM Phase in Heterotic Orbifold  Models \\[10mm]}}
{ Oleg Lebedev \\[6mm]}
\small{Centre for Theoretical Physics, University of Sussex, Brighton BN1
9QJ,~~UK\\[2mm]}
\end{center}

\hrule
\vskip 0.3cm
\begin{minipage}[h]{14.0cm}
\begin{center}
\small{\bf Abstract}\\[3mm]
\end{center}

We consider properties of the CKM phase in the heterotic orbifold models. We find that
at the renormalizable level the CKM phase vanishes identically for the prime orbifolds,
whereas it can be non-zero for some  non-prime orbifolds. In particular, we study in detail 
the $Z_6$-I orbifold which allows for a non-trivial CKM phase and analyze the modular
properties of the corresponding Jarlskog invariant. The CKM phase is shown to vanish 
if the moduli fields are stabilized at ${\rm Im}T_i=\pm 1/2$.

\end{minipage}
\vskip 0.7cm
\hrule
\vskip 1cm

\section{Introduction}

One of the outstanding problems in particle physics is the origin of CP violation.
An attractive possibility is that CP is a good symmetry at the Lagrangian level and
it is the vacuum that breaks it \cite{Lee:1973iz}.
In the context of string theory, this has to be the case since CP is a gauge symmetry 
and thus can be broken only spontaneously \cite{Dine:1992ya}. 
This can be done, for example, by a vacuum
expectation values of the moduli fields \cite{Strominger:1985it}.
In principle, CP can also be violated spontaneously at low energies
in supersymmetric models \cite{Maekawa:1992un},
however this possibility encounters a number of phenomenological difficulties 
\cite{Pomarol:1992bm}. 

Spontaneous CP violation by the VEVs of the moduli fields  in heterotic orbifold models
has recently been studied in Ref.\cite{Bailin:1998xx}. It was found that
order one complex phases in the Yukawa matrices can be produced in this class of
models. However, an important question whether such phases lead to a non-zero
CKM (Cabibbo-Kobayashi-Maskawa) phase was not addressed. This issue will be the focus
of our present work. As we will see, our results are quite different from naive
expectations. In addition, we will study modular properties of the Yukawa couplings
\cite{Lauer:1989ax} and  the  corresponding Jarlskog invariant.

In this letter we consider the possibility of generating a non-zero CKM 
phase at the renormalizable level in heterotic string models with an
orbifold compactification. 
Let us begin by writing the standard quark superpotential as
\begin{equation}
W= Y_{ij}^u H_u Q_{i} U_{j}^c + Y_{ij}^d H_d Q_{i} D_{j}^c \;.
\end{equation}
The CKM phase appears due to the fact that the mass and flavour eigenstates
are generally different, and a complex basis transformation is required
to bring the quark mass matrices into a  diagonal form.
The ``amount'' of CP violation can be quantified in a basis-independent
way via the Jarlskog invariant \cite{Jarlskog:1985ht}:
\begin{eqnarray}
J&=&{\rm Im~\biggl(~ det} \left[ Y^u Y^{u\dagger}, Y^d Y^{d\dagger}  \right]~
\biggr) \nonumber\\
&\propto& (m_t^2-m_u^2)(m_t^2-m_c^2)(m_c^2-m_u^2)
          (m_b^2-m_d^2)(m_b^2-m_s^2)(m_s^2-m_d^2)\nonumber\\
&\times&         {\rm Im}(V_{11}V_{22}V_{12}^* V_{21}^*)\;,
\label{jar}
\end{eqnarray}
where $V_{ij}$ is the CKM matrix. A nonzero value of the Jarlskog invariant
unambiguously indicates the presence of CP violation, i.e. a nontrivial CKM
phase.

\section{Prime Orbifolds }

In the orbifold models the form of the allowed Yukawa couplings is quite
restricted due to various string selection rules 
(for a review see \cite{Bailin:1999nk}). Let us consider
in detail  the $Z_3 \times Z_3$ orbifold \cite{Bailin:1992ak}. 
This orbifold is constructed
via the lattice basis vectors
\begin{equation}
e_i=1 \;\;,\;\;\tilde e_i=e^{2 \pi i /3}\;,
\end{equation}
where $i=1,2,3$ labels the three complex planes,
and the point group generators 
$\theta={\rm diag} (e^{ 2\pi i/3},1,e^{ 4\pi i/3})$
and 
$\omega={\rm diag} (1,e^{ 2\pi i/3},e^{ 4\pi i/3})$
acting on the lattice as
\begin{eqnarray}
\theta : && e_1 \rightarrow \tilde e_1 \;,\;e_2 \rightarrow e_2 \;,\;
            e_3 \rightarrow -e_3-\tilde e_3 \;,\nonumber \\
         && \tilde e_1 \rightarrow -e_1-\tilde e_1 \;,\;
            \tilde e_2 \rightarrow \tilde e_2 \;,\;
            \tilde e_3 \rightarrow e_3 \;, \nonumber \\
\omega:  && e_1 \rightarrow  e_1 \;,\;e_2 \rightarrow \tilde e_2 \;,\;
            e_3 \rightarrow -e_3-\tilde e_3 \;,\nonumber \\
         && \tilde e_1 \rightarrow \tilde e_1 \;,\;
            \tilde e_2 \rightarrow -e_2 - \tilde e_2 \;,\;
            \tilde e_3 \rightarrow e_3 \;.
\end{eqnarray}
In what follows we will only consider twisted matter fields, i.e. fields
whose Yukawa couplings depend on the moduli. This is the only case of interest
since for the untwisted matter fields the discussion becomes trivial. 
Twisted matter fields  belong to the following twisted sectors 
\begin{equation}
\theta \;,\;\theta^2\;,\;\omega\;,\; \omega^2\;,\; \theta\omega^2\;,\;
\theta^2\omega\;,\; \theta\omega
\end{equation}
or $A,\bar A,B,\bar B, C, \bar C, D$, respectively.
In each sector, matter fields are associated with the fixed points or tori
under the corresponding point group element. For instance, let us
consider the $\theta$ (A), $\theta\omega^2$ (C), and $\theta\omega$ (D)
twisted sectors. The A-type fields as well as the C-type fields  are 
associated with  9 fixed tori, whereas the D fields are associated with   
27 fixed points. Explicitly, these fixed tori and fixed points are given by
\begin{eqnarray}
&& f_\theta ={m_1\over 3}(2e_1+\tilde e_1)+{m_3\over 3}(e_3-\tilde e_3)
+z_2~;\; m_{1,3}=0,\pm 1, \nonumber\\
&& f_{\theta\omega^2} ={p_1\over 3}(2e_1+\tilde e_1)+{p_2\over 3}
(e_2-\tilde e_2) +z_3~;\; p_{1,2}=0,\pm 1, \nonumber\\
&& f_{\theta\omega} ={1\over 3}\sum_{i=1}^3 r_i(2e_i+\tilde e_i)\; ;\; r_{i}=0,\pm 1\;,
\label{fixed}
\end{eqnarray}
where $z_{2,3}$ are arbitrary vectors in the second and third complex planes,
respectively.

The trilinear superpotential couplings of twisted fields must obey certain
conditions. First of all, the coupling $f_\alpha f_\beta f_\gamma$ is 
allowed only if the twists $\alpha,\beta,\gamma$ satisfy
\begin{equation}
\alpha\beta\gamma={\bf I}\;,
\end{equation}
which is known as the point group selection rule. Thus only the 
couplings of the type
\begin{equation}
DDD\;,\;\bar ABC\;,\;A\bar B\bar C\;,\; ACD\;,\; B\bar C D\;,\; \bar A
\bar B D 
\end{equation}
are allowed\footnote{These couplings also satisfy the H-momentum
selection rule \cite{Bailin:1999nk}.}.
In addition, the space group selection rule requires
\begin{equation}
({\rm {\bf I}}-\alpha)f_\alpha +({\rm {\bf I}}-\beta)f_\beta+ ({\rm {\bf I}}-\gamma)f_\gamma=0
\end{equation}
up to the addition of $({\rm I}-\alpha)\Lambda_\alpha $,
$({\rm I}-\beta)\Lambda_\beta $ or $({\rm I}-\gamma)\Lambda_\gamma $,
where $\Lambda_i$ are arbitrary lattice vectors. 
This restricts the fixed points that can couple.
For the case of the DDD
coupling $f_{\theta\omega}^{r^1} f_{\theta\omega}^{r^2} 
f_{\theta\omega}^{r^3}$, this selection rule translates into
\begin{equation}
\sum_{J=1}^3 r_i^J =0 \;\;({\rm mod}\;3)\;, i=1,2,3\;\;,
\end{equation}
where $r_i^J$ label the $\theta\omega$ fixed points in the notation 
of Eq.\ref{fixed}. For the ACD coupling 
$f_{\theta} f_{\theta\omega^2} f_{\theta\omega}$ the space selection rule 
 implies
\begin{eqnarray}
&& m_1 +p_1 +r_1=0   \;\;({\rm mod}\;3) \;,\nonumber \\
&& p_2+r_2=0 \;, \nonumber \\
&& m_3+r_3=0 
\label{acd}
\end{eqnarray}
in the notation  of Eq.\ref{fixed}. 

In all cases, if we fix two of the fixed points (tori), 
the third one is determined
unambiguously from the selection rules. This has important implications 
for the structure of the Yukawa matrices. Consider for example the
coupling $ Y_{ij}^u H_u Q_{i} U_{j}^c$. If we assign $H_u$ to a
particular fixed point and $Q_1$ to a different fixed point, there will
be only one fixed point which can couple to this combination. It will
correspond to, for example, $U^c_1$. This implies that the coupling
to $U^c_{2,3}$ vanishes and the resulting Yukawa matrix has a diagonal
form, analogously to the case of the $Z_3$ orbifold \cite{Casas:1990qx}:
\begin{eqnarray}
&& Y^{u,d}=\left( \matrix{a^{u,d} & 0 & 0 \cr
                          0 & b^{u,d} & 0 \cr
                          0 & 0 & c^{u,d}}
\right)\;.
\end{eqnarray} 
Since similar considerations apply to the down-type Yukawa matrix as
well, the Jarlskog invariant vanishes due to $\left[ Y^u,Y^d\right]=0$.
This can also be seen by noting that the complex phases in the Yukawa matrices
can be rotated away by a  redefinition of the right handed fields.
If all the three generations are assigned to the same fixed point,
the resulting Yukawa matrix will have rank 1 leading to degenerate 
eigenvalues and the Jarlskog invariant (Eq.\ref{jar}) vanishes again.
A more interesting structure can be obtained if two generations are assigned
to the same fixed point whereas the third one is assigned to a different fixed 
point\footnote{We assume  to have the freedom to assign a field to a fixed point of
our choice \cite{Casas:1990qx}.}. 
Consider, for instance, the DDD coupling. Let us make the following
$(r_1,r_2,r_3)$ assignment:
\begin{eqnarray}
&& H_u=(0,0,0)\;,\;H_d=(-1,-1,-1)\;,\;Q_{1,3}=(1,1,1)\;,\; Q_2=(1,1,0)\;,\nonumber\\
&& U_{1,3}^c=(-1,-1,-1)\;,\;U_2^c=(-1,-1,0)\;,\;D_{1,2}^c=(0,0,0)\;,\;
D_3^c=(0,0,1)\;.\nonumber
\end{eqnarray}
This leads to the following Yukawa textures
\begin{eqnarray}
&& Y^{u}=\left( \matrix{a & 0 & a \cr
                          0 & b & 0 \cr
                          a & 0 & a}
\right)\;\;,\;\;
  Y^{d}=\left( \matrix{c & c & 0 \cr
                          0 & 0 & d \cr
                          c & c & 0}
\right)\;.
\end{eqnarray} 
For both $Y^u$ and $Y^d$ the complex phases in each column are constant and
therefore can be removed by a phase redefinition of $U_j^c$ and $D_j^c$:
\begin{eqnarray}
&&U_{1,3}^{c'}=e^{i {\rm Arg} (a)}U_{1,3}^c\;,\; 
U_{2}^{c'}=e^{i {\rm Arg} ( b)}U_{2}^c\;,\;
D_{1,2}^{c'} =e^{i {\rm Arg}  (c)}D_{1,2}^c\;,\;
D_{3}^{c'}=e^{i {\rm Arg}  (d)}D_{3}^c\;. \nonumber
\end{eqnarray}
The absence of CP violation in this case can also be seen directly from
Eq.\ref{jar} since $YY^\dagger$ is real for both up- and down-Yukawas.
The same arguments apply to the Yukawa couplings of the other twisted
sectors $ACD , B\bar C D$, etc. in which case the allowed textures
are even more restricted due to fewer allowed parameters involved
(see Eq.\ref{acd}). We thus conclude that the renormalizable couplings
in the $Z_3 \times Z_3$ orbifold model cannot account for the CKM phase
\footnote{We have assumed no flavour mixing in the K\"ahler potential.
Such effects are however insignificant \cite{Casas:1990qx}.}.
These conclusions equally apply to all other prime orbifolds since the
space selection rule for them is of the diagonal type, i.e. for
two given fixed points the third one is selected uniquely.

Let us now comment on other results recently  appeared in the literature.
In Ref.\cite{Giedt:2001es} it was claimed that the CKM phase vanishes in the
$Z_3 \times Z_3$ model if the  (overall) modulus field T gets a VEV at 
the fixed point of the modular group
$\langle T \rangle =\exp(i\pi /6) $.
The argument is that the Yukawa couplings in this case can be expressed as
\begin{equation} 
Y^u_{ij}= \vert Y^u_{ij} \vert~ e^{i \alpha^u_i} \;\;,\;\;
Y^d_{ij}= \vert Y^d_{ij} \vert~ e^{i \alpha^d_i}
\end{equation}
with $\alpha_1^u=0~,~\alpha_{2,3}^u=-\pi/3~,~ \alpha_2^d=0~,~
\alpha^d_{1,3}=-\pi/3$. In other words, the complex phases are constant
in each row and are either 0 or $-\pi/3$. 
It was claimed that for this particular set of the phases the CKM phase
vanishes regardless of $\vert Y^{u,d}_{ij} \vert$.
However, numerous counterexamples to this statement can be found.
For example, it is easy to check numerically that for $\vert Y^{u}_{ij} \vert
=i+j $ and $\vert Y^{d}_{ij} \vert =i+2j $ the Jarlskog determinant does not
vanish and thus CP is violated.
The technical flaw in the considerations of Ref.\cite{Giedt:2001es} was
to assume that the matrix ${\rm diag}(e^{i\pi/3},e^{-i\pi/3},1)$
commutes with arbitrary orthogonal matrices.
From what we have seen above, it is clear that the CKM phase vanishes
due to the restricted flavour structure of the Yukawa matrices
rather than a specific phase assignment. It is also clear that the modular
group fixed point  $\exp(i\pi /6) $ is not special in this respect and
the CKM phase vanishes for any other $\langle T \rangle$ as well.

The situation may change if higher order operators are taken into account.
These operators are required anyway if we are to produce
the observed fermion mass hierarchy and mixings \cite{Casas:1992zt}. To this end, the $Z_3$
orbifold seems most promising since it allows for 9 deformation parameters
(as opposed to 3 in the $Z_3 \times Z_3$ case), which can play a
significant role in fitting the fermion masses.

\section{$Z_6$-I and Other Non-Prime Orbifolds}

Let us now turn to the discussion of the non-prime orbifolds. They are essentially
different from the prime orbifolds in that the space group selection rule is 
non-diagonal and for given two fixed  points the third one is not selected uniquely.
This entails a much broader variety of allowed Yukawa textures and, as we will see below,
a possibility of generating a non-trivial CKM phase at the renormalizable level.
Let us consider the $Z_6$-I orbifold as an example (for further details see 
Ref.\cite{Casas:1993ac}).

The $Z_6$-I orbifold is formed by the $G_2 \times G_2 \times SU(3)$
lattice
\begin{eqnarray}
&& e_i=1\;,\; \tilde e_i= \sqrt{3}~ e^{5 \pi i/6}\;\;,\;\;i=1,2\;,\nonumber\\
&& e_3=1\;,\; \tilde e_3=  e^{2 \pi i/3}\;,
\end{eqnarray}
and the twist $\theta={\rm diag} (e^{ i\pi /3},e^{ i\pi /3},e^{ -2\pi i/3})$
acting on the lattice as
\begin{eqnarray}
&& \theta e_1=-e_1-\tilde e_1  \;\;,\;\; \theta \tilde e_1=3e_1+2\tilde e_1\;,\nonumber\\
&& \theta e_2=-e_2-\tilde e_2  \;\;,\;\; \theta \tilde e_2=3e_2+2\tilde e_2\;,\nonumber\\
&& \theta e_3=\tilde e_3  \;\;,\;\; \theta \tilde e_3=-e_3-\tilde e_3\;.
\end{eqnarray}
The orbifold fixed points  fall into the three categories: $\theta$, $\theta^2$, 
and $\theta^3$. Contrary to the case of the prime orbifolds, a point fixed under
$\theta^2$ or $\theta^3$ is not necessarily fixed under $\theta$. 
However, physical states must be eigenstates of the twist $\theta$. As a result,
physical states correspond to the conjugation classes of the fixed points under
$\theta$ rather than the fixed points themselves \cite{Kobayashi:1990mc}. 
That is to say, two fixed points belong to the same conjugation class if they can 
be connected by a $\theta$ ($\theta^2$) transformation. Formally, if $f_k$
is a $\theta^k$ fixed point and $l$ is the smallest number such that 
$f_k$ is fixed under $\theta^l$, the physical states are expressed as 
\begin{equation}
\vert f_k \rangle + e^{-i\gamma} \vert \theta f_k \rangle +...+
e^{-i (l-1) \gamma} \vert \theta^{l-1} f_k \rangle \;,
\label{conjug}
\end{equation} 
where $\gamma=2\pi p/l$ and $p=1,2,..,l$. It can be easily verified that such states are 
indeed eigenstates of $\theta$.
For the $Z_6$-I orbifold there are 3 conjugation classes (3 fixed points) in the
$\theta$ sector, 15 conjugation classes (27 fixed points) in the
$\theta^2$ sector, and 6 conjugation classes (16 fixed tori) in the
$\theta^3$ sector. 

Let us consider the $Z_6$-I orbifold fixed points and their couplings in more detail. 
In terms  of the $G_2 \times G_2$ lattice basis, the fixed points can be written as 
(a tensor product with the three SU(3) lattice $Z_3$ fixed points or
a fixed 2-torus for the $\theta^3$ sector is understood)
\begin{eqnarray}
 \theta-{\rm sector:}\;\;\;
&&g_1^{(0)} \otimes g_1^{(0)}\;,\nonumber\\
 \theta^2-{\rm sector:}\;\;\;
&&g_2^{(i)} \otimes g_2^{(j)}\;,\nonumber\\
 \theta^3-{\rm sector:}\;\;\;
&&g_3^{(i)} \otimes g_3^{(j)}\;,
\end{eqnarray}
where
\begin{eqnarray}
&& g_1^{(0)}=(0,0)\;\;, \nonumber\\
&& g_2^{(i)}=\biggl[ (0,0),\left( 0,{1\over 3} \right),\left( 0,{2\over 3} \right) 
\biggr] \;\;, \nonumber\\
&& g_3^{(i)}=\biggl[ (0,0),\left( 0,{1\over 2} \right),\left( {1\over 2},0 \right) , 
\left( {1\over 2},{1\over 2} \right) \biggr]\;. 
\end{eqnarray}
The point group selection rule and the H-momentum conservation allow only the
Yukawa couplings of the form 
\begin{equation}
\theta \theta^2 \theta^3 \;\;\;,\;\;\; \theta^2 \theta^2 \theta^2\;.
\end{equation}
The  space group selection rule for the coupling $\theta \theta^2 \theta^3$
requires \cite{Casas:1993ac}
\begin{equation}
f_1+({\bf I}+\theta)f_2 -({\bf I} +\theta +\theta^2 )f_3 \in \Lambda \;,
\end{equation}
where $f_{1,2,3}$ belong to the $ \theta, \theta^2, \theta^3$
twisted sectors, respectively, and $\Lambda$ denotes the orbifold lattice.
It can be easily verified that this condition imposes no restriction on the
$G_2 \times G_2$ components of the fixed points and requires that the
SU(3) components of $f_1$ and $f_2$ be equal:
\begin{equation}
f_1\Bigl\vert_3=f_2\Bigl\vert_3\;.
\label{space}
\end{equation}
\begin{table}
\begin{center}
\begin{tabular}{|c||c|c|}
\hline
  {\rm field}   & $G_2 \times G_2$  {\rm fixed point} & $l$  \\
\hline
\hline
$H_{1,2}$ & $(0,0)\otimes (0,0) $ & 1 \\
$Q_1$ & $(0,0)\otimes (0,0) $ & 1 \\
$Q_2$ & $\left( 0,{1\over3} \right) \otimes \left( 0,{1\over3} \right) $ & 2 \\
$Q_3$ & $\left( 0,{1\over3} \right) \otimes \left( 0,0 \right) $ & 2 \\
$U_1$ & $(0,0)\otimes (0,0) $ & 1 \\
$U_2$ & $\left( 0,{1\over2} \right) \otimes \left( 0,{1\over2} \right) $ & 3 \\
$U_3$ & $\left( {1\over2},{1\over2} \right) \otimes \left( 0,{1\over2} \right) $ & 3 \\
$D_1$ & $\left( 0,{1\over2} \right) \otimes \left( 0, 0 \right) $ & 3 \\
$D_2$ & $\left( 0,0 \right) \otimes \left( 0,{1\over2} \right) $ & 3 \\
$D_3$ & $(0,0)\otimes (0,0) $ & 1 \\
\hline
\end{tabular}
\end{center}
\caption{$Z_6$-I fixed point assignment for the observable fields
in the $(e_1,\tilde e_1, e_2, \tilde e_2)$ basis. $l$ indicates a number
of the fixed points in the corresponding conjugation class. }
\label{z6}
\end{table}
Thus, there are numerous combinations of the fixed points which can couple and
various Yukawa textures can be produced. Suppose $H_{1,2}$ belong
to the $\theta$-sector, $Q_i$ to the $\theta^2$-sector, and 
$U_i , D_i$ to the $\theta^3$-sector, and associate observable fields with
the fixed points (tori) as shown in Table \ref{z6}. As before, we omit
the $SU(3)$ lattice components which are fixed by Eq.(\ref{space}).
In Table \ref{z6}, we also present the number of the fixed points $l$ in each
conjugation class (see Eq.\ref{conjug}). If $l$ is greater than one, we associate
a physical field with a symmetric combination of the elements of the conjugation 
class (i.e. $\gamma=2\pi$ in Eq.\ref{conjug}), since only symmetric combinations
enter the coupling $\theta \theta^2 \theta^3$ \cite{Casas:1993ac}.

The corresponding $f_1 f_2 f_3$ Yukawa couplings are expressed as \cite{Casas:1993ac}
\begin{eqnarray}
&& Y_{\theta \theta^2 \theta^3}=
N \sqrt{l_2 l_3} \sum_{\stackrel{\rightarrow}{u} \in Z^4}
{\rm exp} \biggl[ -4\pi  \left( 
\stackrel{\rightarrow}{f_{23}} + \stackrel{\rightarrow}{u} \right)^T
M \left( 
\stackrel{\rightarrow}{f_{23}} + \stackrel{\rightarrow}{u} \right)
\biggr]\;,
\end{eqnarray}
where $\stackrel{\rightarrow}{f_{23}}$ represents the $G_2\times G_2$
projection of $f_2-f_3$ in the basis $(e_1,\tilde e_1, e_2, \tilde e_2)$,
$\stackrel{\rightarrow}{u}$ is a four-dimensional vector with integer
components, $N$ is a normalization factor,
 and the matrix $M$ is
given by
\begin{eqnarray}
&& M=\left( \matrix{T_1 & -{3\over2}T_1 & 0 & 0 \cr
                    -{3\over2}T_1 & 3T_1 & 0 & 0 \cr
                    0 & 0 & T_2 & -{3\over2}T_2 \cr
                    0 & 0 & -{3\over2}T_2 & 3T_2} \right)\;.
\end{eqnarray}
For simplicity we have assumed no lattice deformations and have used the
following relation between the moduli fields $T_i$ and the compactification radii $R_i$
\cite{Bailin:1999nk}
\begin{equation}
{\rm Re}T_i= {1\over 4}~ {2 \sqrt{{\rm det}~g_i}\over (2\pi)^2}= {\sqrt{3} \over 16 \pi^2} R_i^2\;,
\end{equation}
where $g_{ab}=e_a \cdot e_b$ and the factor 1/4 appears due to a difference in the 
definitions of Refs.\cite{Bailin:1999nk} and \cite{Casas:1993ac}.
Note that only the 
$G_2\times G_2$ lattice components of the fixed points affect the Yukawa couplings.
This occurs due to the fact that $\theta^3$ leaves the third plane invariant 
and thus the third plane does not contribute to the classical action.

We find that the Yukawa matrices corresponding to the assignment in Table \ref{z6}
lead to  a non-zero Jarlskog invariant and thus produce a CKM phase. 
In the next section
we will study the numerical behaviour and modular properties of the  Jarlskog invariant.

So far we have concentrated on the coupling of the type $\theta \theta^2 \theta^3$.
In the $Z_6$-I orbifold, we can also have a  $\theta^2 \theta^2 \theta^2$ coupling.
In this case, however, the analysis  is trivial since
the corresponding space group selection rule is diagonal \cite{Casas:1993ac}  and the 
CKM phase vanishes. We find that even for the non-prime orbifolds the CKM phase often
vanishes since the space group selection rule is typically quite restrictive although  
not diagonal. For instance, we have analyzed the $Z_4$ orbifold with the $\left[SO(4)\right]^3$
lattice and have not found a non-trivial CKM phase. A detailed investigation of all orbifolds
allowing for a non-zero CKM phase will be presented elsewhere.

\section{Jarlskog Invariant and Modular Transformations }

Let us analyze the properties of the Jarlskog determinant for the $Z_6$-I orbifold
with the field assignment of Table 1  under a modular
transformation 
\begin{equation}
T_i \rightarrow { a T_i - ib \over ic T_i +d}\;,
\end{equation}
where $ad-bc=1$. The $SL(2,Z)$ group of such transformations is generated by two generators,
$T_i \rightarrow T_i+i$ and $T_i \rightarrow 1/T_i$, and for our purposes it suffices to consider  
 these two transformations only.   

Let us first consider the shift
$T_{1,2} \rightarrow T_{1,2}+i$. We find  that the Yukawa matrices
transform under the axionic shift as
\begin{eqnarray}
&&Y^u \rightarrow \left( \matrix {1& 0 & 0 \cr
                             0 & e^{-2\pi i/3} & 0 \cr
                             0 & 0 & e^{2\pi i/3}}     \right)~ Y^u  \;,\nonumber\\
&&Y^d \rightarrow \left( \matrix {1& 0 & 0 \cr
                             0 & e^{-2\pi i/3} & 0 \cr
                             0 & 0 & e^{2\pi i/3}}     \right)~ Y^d ~  
\left( \matrix {e^{i\pi}& 0 & 0 \cr
                             0 & e^{i\pi} & 0 \cr
                             0 & 0 & 1}     \right) \;.
\label{transform}
\end{eqnarray}
Note that the phase matrices multiplying the Yukawas from the left
are the same for  the up and down sectors. As a result, such phases can be
absorbed into the definition of the quark doublets and down-type singlets:
$Q_i \rightarrow e^{i\alpha_i} Q_i$ and   $D_i^c \rightarrow e^{i\beta_i}D_i^c$,
where $\alpha_i=\left( 0, 2\pi/3, -2\pi/3\right)$ and 
$\beta_i=\left( -\pi , -\pi, 0 \right)$. Clearly, the Jarlskog 
determinant stays invariant under this transformation, as it should.
The reason for the above  transformation property (Eq.\ref{transform}) is the ``phase
factorization'', i.e. for given two fixed points $f_2$ and $f_3$ we have
\begin{equation}
Y\left( f_2-f_3; T_{1,2}+i\right)=Y\left( f_2-f_3; T_{1,2}\right) e^{i \phi(f_2)} e^{i \phi(f_3)}\;.
\label{factor}
\end{equation} 
This can be seen as follows. Consider for simplicity the second complex plane only.
Under $T_2\rightarrow T_2+i$ the Yukawa matrix will pick up a phase
\begin{equation}
\exp \left[ -4\pi i\left( f_{23}^{(1)2}-3f_{23}^{(1)} f_{23}^{(2)}+3f_{23}^{(2)2} \right)
  \right]\;, 
\end{equation}
where the superscripts $(1),(2)$ refer to the coordinates  in the lattice basis 
($e_2,\tilde e_2$). Substituting $f_{23}\equiv f_2-f_3$ and recalling
that $f_2=(0,l/3)$, $f_3=(m/2,n/2)$ with integer $l,m,n$, we readily see that the cross terms
$f_2^{(i)} f_3^{(j)}$ disappear (up to $2\pi i$) and Eq.\ref{factor} is satisfied. 
The phase factorization property implies that under the axionic shift
\begin{equation}
Y^u_{ij} \rightarrow Y^u_{ij}~ e^{i(\alpha_i+\beta^u_j)}\;\;,\;\;
Y^d_{ij} \rightarrow Y^d_{ij}~ e^{i(\alpha_i+\beta^d_j)}\;.
\end{equation}
Such phase factors can always be absorbed in the redefinition of the fields
and the Jarlskog determinant remains invariant (Fig.1).  

Let us now consider the duality transformation $T_2 \rightarrow 1/T_2$ in the second
complex plane. For convenience,
we introduce auxiliary  quantities $\chi^{ij}$ defined as
\begin{eqnarray}
&& \chi^{ij}=
\sum_{\stackrel{ }{\tilde u} \in Z^2}
{\rm exp} \biggl[ -4\pi T_2  \left( 
\stackrel{ }{\tilde f_{23}^{ij}} + \stackrel{ }{\tilde u} \right)^T
{\bf {\rm m}} \left( 
\stackrel{ }{\tilde f_{23}^{ij}} + \stackrel{ }{\tilde u} \right)
\biggr]\;,
\end{eqnarray}
with 
\begin{eqnarray}
&& {\bf {\rm m}}=\left( \matrix{1 & -{3\over2}  \cr
                               -{3\over2} & 3   \cr
                                } \right)\;.
\end{eqnarray}
Here the tilded quantities refer to the projections on the second complex plane and $ij$ labels
all possible $\tilde f_{23}$ ($i$ enumerates $\tilde f_2$ and 
$j$ enumerates  $\tilde f_3$). Since 
$\tilde f_2=\left[ (0,0),\left( 0, 1/3 \right),\left( 0, 2/3 \right) \right]$ and
$\tilde f_3=\left[ (0,0),\left( 0,{1/2} \right),\left( 1/2,0 \right) , 
\left( 1/2, 1/2 \right) \right]$, 
there are 12 different $\tilde f_{23}$. However, it can be shown that only
four of them produce different $\chi$'s. Indeed, the couplings $\chi$ are the same for each 
conjugation class and their representatives can be chosen as  
$\tilde f_2=\left[ (0,0),\left( 0, 1/3 \right) \right]$ and 
$\tilde f_3=\left[ (0,0),\left( 0, 1/2 \right) \right]$. The resulting inequivalent $\chi$'s
are generated by 
\begin{equation}
\tilde f_{23}=\biggl[ (0,0),\left( 0, 1/2 \right),
\left( 0, 1/3 \right),\left( 0, 1/6 \right) \biggr]\;. 
\end{equation}
We will label the corresponding $\chi$'s as $(\chi^1,\chi^2,\chi^3,\chi^6)$ referring to the number
of $\tilde f_{23}$'s producing the same $\chi$.
Using the Poisson resummation formula
\begin{eqnarray}
&&  \sum_{ m \in Z^d}
{\rm exp} \biggl[ -\pi   \left( 
\stackrel{ }{ m } + \stackrel{ }{\delta} \right)^T
{\bf {\rm A}} \left( 
\stackrel{ }{m} + \stackrel{ }{\delta} \right)
\biggr] \nonumber\\
&& =
{1\over \sqrt{{\rm det~A} }} \sum_{ m \in Z^d}
{\rm exp} \biggl[ -\pi  m^T
{\bf {\rm A^{-1}}}m -2\pi i ~\delta^T m  
\biggr]
\end{eqnarray}
and rearranging the sums, we find the following transformation properties for the $\chi$'s:
\begin{eqnarray}
&& \left( \matrix{ \chi^1 \cr
                   \chi^2 \cr
                   \chi^3 \cr
                   \chi^6 \cr 
}  \right) \left( {1\over T_2} \right)= {T_2 \over 2 \sqrt{3}}  \left( \matrix{1 & 2 & 3 & 6   \cr
                                   1 & -1 & 3 & -3 \cr
                                   1 & 2 & -1 & -2 \cr
                                   1 & -1 & -1 & 1} \right)
\left( \matrix{    \chi^1 \cr
                   \chi^2 \cr
                   \chi^3 \cr
                   \chi^6 \cr 
}  \right) (T_2)\;.
\end{eqnarray}
In terms of the original 12 dimensional basis $\chi^{ij}$, this transformation is unitary (times $T_2$).
It is given by  a tensor product $A \otimes B$ with
\begin{eqnarray}
&&A={1\over \sqrt{3}} \left( \matrix{1 & 1 & 1    \cr
                                   1 & e^{-2\pi i/3} & e^{2\pi i/3}  \cr
                                   1 & e^{2\pi i/3} & e^{-2\pi i/3}  \cr
                                   } \right) \;,\;
B={1\over 2} \left( \matrix{1 & 1 & 1 & 1   \cr
                                   1 & -1 & 1 & -1  \cr
                                   1 & -1 & -1 & 1  \cr
                                   1 & 1 & -1 & -1 
                            } \right),
\end{eqnarray} 
such that 
\begin{equation}
\chi^{ij}\left( {1\over T_2} \right)= T_2~ A^{ii'}B^{jj'}\chi^{i'j'}\;.
\label{unit}
\end{equation}
Since in the absence of the lattice deformations the Yukawa couplings can be factorized 
$Y_{ab} \propto \chi^{ij}(T_1) \chi^{i'j'}(T_2) $, we have similar transformation
properties for the Yukawas.

It is well known that the most general transformation of the Yukawa matrices preserving
the Jarlskog determinant is 
\begin{equation}  
Y^u \rightarrow U_L Y^u U_R^{u \dagger} \;\;,\;\; Y^d \rightarrow U_L Y^d U_R^{d \dagger}\;,
\end{equation}
where $U_L, U_R^{u,d}$ are $3\times3$ unitary matrices. The duality transformation of
Eq.\ref{unit} does not belong to  this class. Indeed, we have 
$\chi \rightarrow A \chi B^\dagger $ (up to a rescaling) where $B$ is a $4\times4$ unitary matrix.
The corresponding action on the quark generations is not unitary. This stems from the fact
that the duality transformation is unitary $only$  when acting  on the fields associated with all
of the fixed points. Since the number of the fixed points is larger than three, the corresponding
action on the three quark generations is not unitary\footnote{$B$ does not contain   unitary
blocks of a lower dimension. The same applies to the $\theta$-eigenstate basis (\ref{conjug}).}. 
Of course, the entire superpotential $Y_{abc} \phi^a \phi^b \phi^c$ with the sum  taken
over all of the fixed points is modular covariant. However, its subset describing the 
Standard Model interactions is not. The pieces necessary to restore the modular covariance are 
associated with heavy matter fields and are ``decoupled'' from the low energy theory.
As a result, the Jarlskog determinant $does$ $not$ transform covariantly under the duality
transformation (unless it is zero). Below we will also demonstrate it numerically.

The axionic shift invariance allows us to derive an important property of the CKM phase.
The CKM phase has to vanish if the moduli fields are stabilized at ${\rm Im}T_i=\pm 1/2$,
which includes the fixed points of the modular group ${\rm exp}(\pm i \pi /6)$. Indeed,
since $T_i^*=T_i \pm i$ the Jarlskog invariant satisfies 
\begin{equation}
J \left[  Y(T_i)\right]=-J \left[  Y^* (T_i) \right]=-J \left[  Y(T_i^*)\right]=
-J \left[  Y(T_i)\right]\;,
\end{equation}
where we have used the fact that the Yukawa couplings are holomorphic functions of the moduli fields.
As a result, the CKM phase vanishes. Note that if the Jarlskog determinant transformed
$covariantly$ under the duality, the CKM phase
would have to vanish on the unit circle by the same argument. This is however not the case
as illustarted in Fig.2. 
This fact was not taken into account in Ref.\cite{Dent:2000ki} which resulted in a misleading
conclusion.

The $\vert T \vert $-dependence of the Jarlskog invariant is shown in Fig.3.
For ${\rm Re}T \sim  1$, $J(T)$ falls off exponentially 
(which accounts for the difference in the scales of Figs.1 and 2). The numerical value of 
$J(T)$ is not important since we are not attempting to produce the correct quark masses and 
mixings. Non-renormalizable operators must be included to produce a more realistic picture
\cite{Casas:1990qx},\cite{Casas:1992zt},\cite{Kobayashi:1995py}.
 
It should be noted that we have used unnormalized Yukawa couplings throughout the paper.
The Yukawa couplings for the properly normalized fields are obtained by the rescaling 
$ Y_{abc}\rightarrow Y_{abc} {\hat W^*}/\vert {\hat W} \vert e^{\hat K/2}
(K_a K_b K_c)^{-1/2}$ \cite{Brignole:1997dp}. 
Due to the modular weight sum rule of Ref.\cite{Ibanez:1992hc}, this amounts
to a multiplication of $Y_{abc}$ by $\sqrt{T_i+\bar T_i}$
(up to a phase), which makes it a weight zero quantity. 
The Jarlskog
invariant is insensitive to a phase redefinition and  is simply rescaled by $(T_i+\bar T_i)^6$. 
Since we are concerned with qualitative 
behaviour of the CKM phase, these rescaling effects are not important for the 
present study.

The above results can be generalized to higher non-prime order orbifolds. 
Indeed, the phase factorization
property of Eq.\ref{factor} is quite general, so is the axionic shift symmetry. As a result, 
the CKM phase has to vanish for ${\rm Im}T_i=\pm 1/2$. On the other hand, 
generally there is no duality symmetry in the Standard Model sector
(unless $J=0$) as there are even more fixed points than in the $Z_6$  case.
Thus it is possible to produce a non-trivial CKM phase for $T_i$ on the unit circle
(apart from the fixed points).

{\bf Acknowledgements}. The author is indebted to David Bailin for valuable discussions.

\newpage

\thispagestyle{empty}

\begin{figure}[p]
\begin{center}
\begin{tabular}{c}
\epsfig{file=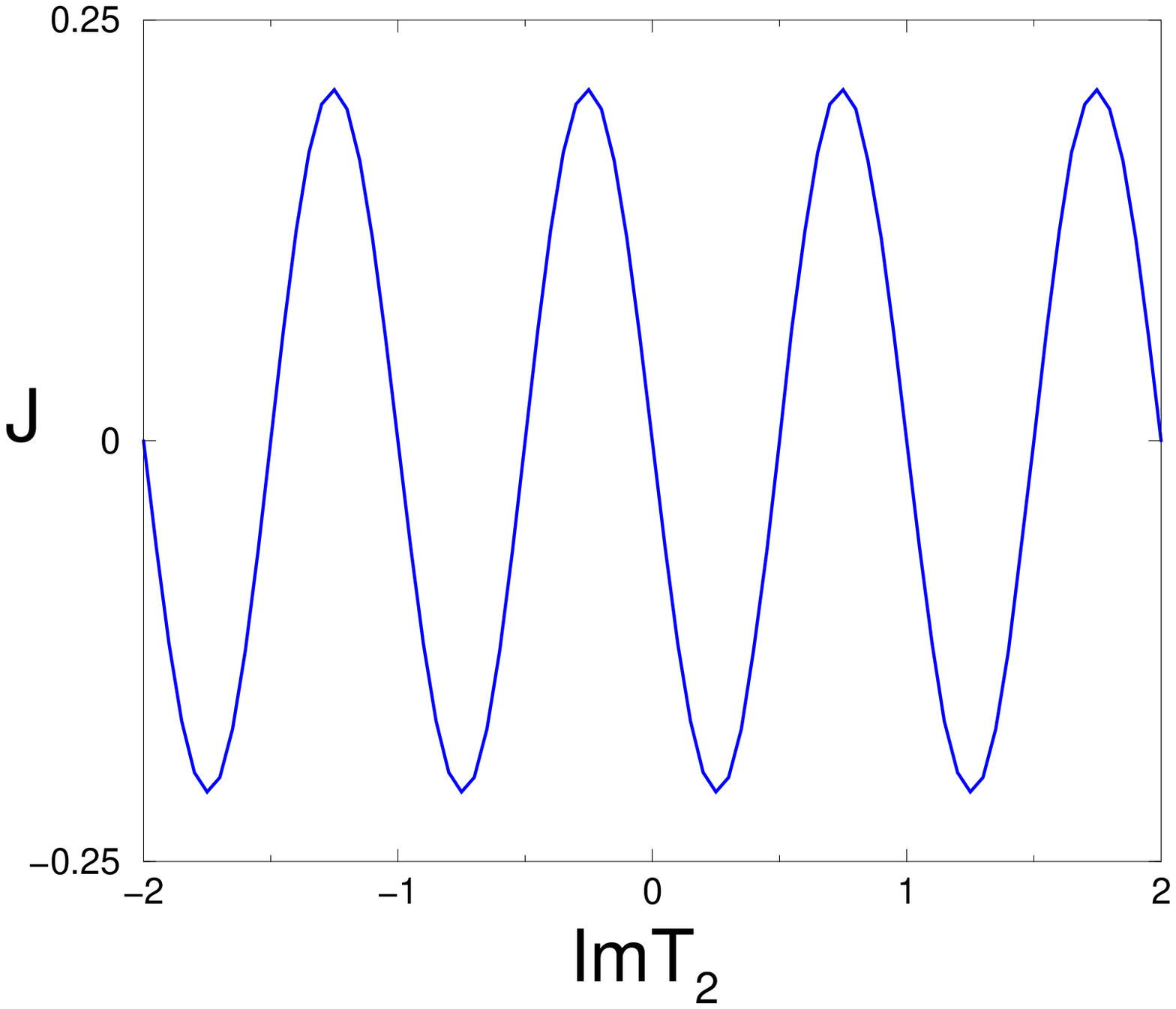, width=7.4cm, height=5.8cm}\\  
\end{tabular}
\end{center}
\caption{Jarlskog invariant as a function of  ${\rm Im} T_2$
for $T_1=0.3$ and  ${\rm Re} T_2=1$.
 }
\label{axion}
\end{figure}
\begin{figure}[ht]
\begin{center}
\begin{tabular}{c}
\epsfig{file=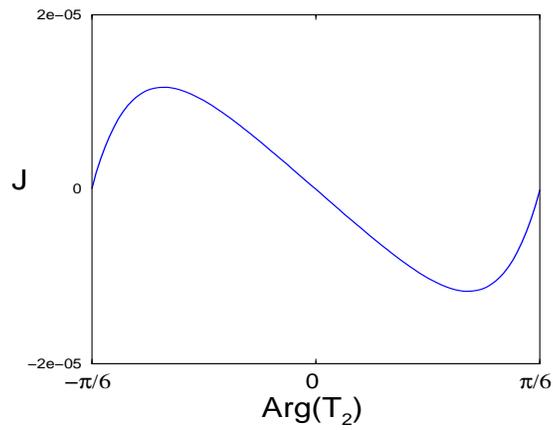, width=7.4cm, height=5.8cm}\\  
\end{tabular}
\end{center}
\caption{Jarlskog invariant as a function of  ${\rm Arg}( T_2 )$
on the unit circle ($T_1=1,\vert T_2 \vert=1$). 
 }
\label{circle}
\end{figure}
\begin{figure}[ht]
\begin{center}
\begin{tabular}{c}
\epsfig{file=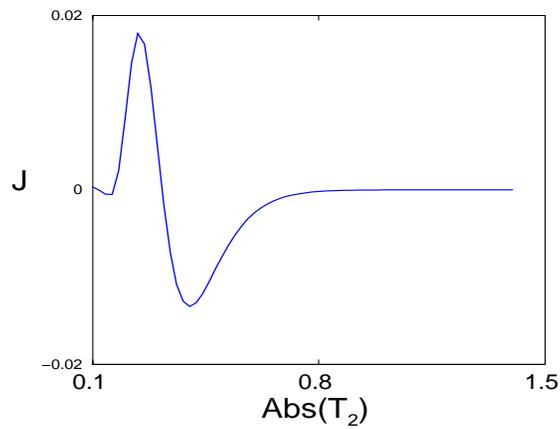, width=7.4cm, height=5.8cm}\\  
\end{tabular}
\end{center}
\caption{Jarlskog invariant as a function of  $\vert T_2 \vert $
for  $T_1=1, {\rm Arg}( T_2) = \pi/6$. 
 }
\label{abs}
\end{figure}
\clearpage
 
\end{document}